\documentclass[letterpaper,twocolumn,showpacs,prb,superscriptaddress,email]{revtex4}
\pdfoutput=1
\usepackage{natbib}
\usepackage[pdftex]{graphicx}
\usepackage{amssymb}
\usepackage{amsmath,amsfonts,latexsym}
\usepackage{array,tabularx,color}
\usepackage{dcolumn}               
\usepackage[normalem]{ulem}
\usepackage{comment}
\usepackage{bm}
\usepackage[latin1]{inputenc}
\usepackage{color}

\begin{document}
\title{Merging of vortices and antivortices in polariton superfluids}

\author{E. Cancellieri} \affiliation{Laboratoire Kastler Brossel, Universit\'e Pierre et Marie Curie, Ecole Normale Sup\'erieure et CNRS, UPMC case 74, 4 place Jussieu, 75005 Paris, France} \affiliation{Department of Physics and Astronomy, University of Sheffield, Sheffield, S3 7RH, United Kingdom}

\author{T. Boulier} \affiliation{Laboratoire Kastler Brossel, Universit\'e Pierre et Marie Curie, Ecole Normale Sup\'erieure et CNRS, UPMC case 74, 4 place Jussieu, 75005 Paris, France}

\author{R. Hivet} \affiliation{Laboratoire Kastler Brossel, Universit\'e Pierre et Marie Curie, Ecole Normale Sup\'erieure et CNRS, UPMC case 74, 4 place Jussieu, 75005 Paris, France}

\author{D. Ballarini} \affiliation{NNL, Istituto Nanoscienze - CNR, Via Arnesano, 73100 Lecce, Italy} \affiliation{CNB@UniLe, Istituto Italiano di Tecnologia, via Barsanti, 73100 Arnesano (Lecce), Italy}

\author{D. Sanvitto} \affiliation{NNL, Istituto Nanoscienze - CNR, Via Arnesano, 73100 Lecce, Italy} \affiliation{CNB@UniLe, Istituto Italiano di Tecnologia, via Barsanti, 73100 Arnesano (Lecce), Italy}

\author{M. H. Szymanska} \affiliation{Department of Physics and Astronomy, University College London, Gower Street, London, WC1E 6BT, UK}

\author{C. Ciuti} \affiliation{Laboratoire Mat\'eriaux et Ph\'enom\'enes Quantiques, UMR 7162, Universit\'e Paris Diderot-Paris 7 et CNRS, 75013 Paris, France}

\author{E. Giacobino} \affiliation{Laboratoire Kastler Brossel, Universit\'e Pierre et Marie Curie, Ecole Normale Sup\'erieure et CNRS, UPMC case 74, 4 place Jussieu, 75005 Paris, France}

\author{A. Bramati} \affiliation{Laboratoire Kastler Brossel, Universit\'e Pierre et Marie Curie, Ecole Normale Sup\'erieure et CNRS, UPMC case 74, 4 place Jussieu, 75005 Paris, France}

\begin{abstract}
Quantised vortices are remarkable manifestations on a macroscopic
scale of the coherent nature of quantum fluids, and the study of
their properties is of fundamental importance for the understanding
of this peculiar state of matter. Cavity-polaritons, due to their
double light-matter nature, offer a unique controllable environment
 to investigate these properties. In this work we theoretically investigate
 the possibility to deterministically achieve the annihilation of a vortex
 with an antivortex through the increase of the polariton density 
in the region surrounding the vortices. Moreover we demonstrate
that by means of this mechanism an array of vortex-antivortex
pairs can be completely washed out.
\end{abstract}

\maketitle

\section{Introduction}
Topological excitations such as quantised vortices, characterised by a phase
winding from 0 to $2\pi m$ (with $m$ an integer number) around a
vortex core, have been extensively studied in several systems such
as non-linear optical systems \cite{desyatnikov_optical_2005},
superconductors \cite{essmann_direct_1967}, superfluid $^4$He
\cite{yarmchuk_observation_1979}, vertical-cavity surface-emitting
lasers \cite{scheuer_optical_1999}, and more recently in cold atoms
\cite{madison_vortex_2000,denschlag_generating_2000,khaykovich_formation_2002}.
Finally, in recent years, the study of vortices and vortex lattices
has attracted much attention also in the field of coherent
cavity-polariton fluids. First because, being intrinsically
out-of-equilibrium, they constitute a novel system to study
Bose-Einstein condensation phenomena, and secondly because, due to
their light-matter nature, polaritonic systems are
fully controllable by optical techniques and
therefore allow very detailed studies of
quantum turbulence.

In particular, in the context of cavity-polariton systems, it has
been shown that stable vortices and half-vortices
\cite{lagoudakis_quantized_2008,lagoudakis_observation_2009,flayac_topological_2010},
as well as single vortex-antivortex (V-AV) pairs
\cite{roumpos_single_2011,nardin_hydrodynamic_2011,tosi_onset_2011,sanvitto_all-optical_2011}
can be generated. The formation of
lattices of vortices and of vortex-antivortex pairs has also been
theoretically and experimentally studied in several different
configurations: in the optical parametrical oscillator configuration
\cite{gorbach_vortex_2010}, in non-resonantly generated condensates
\cite{keeling_spontaneous_2008,tosi_geometrically_2012,cristofolini_optical_2013},
and in the case of patterns induced by metallic deposition on the
surface of the cavity \cite{kusudo_stochastic_2012}. However,
 the mechanisms lying beneath vortex-antivortex annihilation and vortex-vortex
interaction are still not fully understood. This is related
to the high degree of control needed to study such interactions.

For example, in the first experimental observations of
vortex-antivortex lattices
\cite{tosi_geometrically_2012,cristofolini_optical_2013,kusudo_stochastic_2012}
the formation and the properties of the array were
only partially controllable due to either the presence of a strong
exciton reservoir, that influences
the position of the formed vortex array and its disappearance, or
due to the fact that the formed array depends on
the structure of the metallic depositions over the cavity surface.
To achieve a higher degree of control resonant pumping schemes have
been proposed \cite{liew_generation_2008}. By using
 masks in the pumping beam, the formation of
vortex-antivortex arrays with controllable shape and vortex
distribution was achieved \cite{hivet_2014}
and the evolution of the arrays was studied in correlation
with the local onset of the superfluid regime. 
However, in this study the case
of high polariton densities was not experimentally achievable
since the masks used to generate the vortex array were
blocking most of the laser power, and therefore was not
theoretically investigated.

In this work we theoretically study the annihilation of a
vortex with an antivortex when injecting polaritons resonantly,
in a broad range of polariton densities. In our model we assume an
excitation with four coherent laser beams resonant with the
lower polariton branch similarly to \cite{tosi_geometrically_2012} and
\cite{cristofolini_optical_2013} where, however, polaritons were
injected non-resonantly. The four pump spots are supposed to have
the same energy and k-vector modulus and to generate polaritons
propagating toward the centre of a common area. The advantage of
our model is that in the resonant configuration it is well known that
there is no exciton reservoir and that, since the entire pump intensity is
used to inject the coherent fluid, high polariton densities can be achieved.
Moreover, since the pumps set the momentum and the density of the
injected fluid we can directly correlate the vortex-antivortex annihilation
and the washing out of an array of vortices with the increase of the regions
where the fluid is subsonic and with the change in the polariton flow
that comes with it.

The manuscript is structured as follows: In Sec. \ref{Model} we introduce the
theoretical model used to simulate the system, describe more in details the
setup we have in mind and introduce the {\it generalized local speed of sound} that will
be needed for the analysis of the results. In Sec. \ref{Results} we demonstrate the
annihilation of V-AV pairs and correlate this annihilation with the change in
the polariton flow caused by the widening of the regions where the fluid is subsonic.
Finally in In Sec. \ref{Conclusion} we draw some conclusions and give prospects
for future developments of the work.

\section{Model}\label{Model}
A standard way to model the dynamics of resonantly-driven polaritons
in a planar microcavity is to use a Gross-Pitaevskii (GP) equation
\cite{carusotto_quantum_2013} for coupled cavity and exciton fields
($\Psi_C$ and $\Psi_X$) generalized to include the effects of the
resonant pumping and decay ($\hbar=1$):

\begin{equation*}
\partial_t \begin{pmatrix} \Psi_X \\ \Psi_C \end{pmatrix}
=\begin{pmatrix} 0 \\ F \end{pmatrix} + \left[ H_0 + \begin{pmatrix} g_X|\Psi_X|^2 & 0 \\ 0 & V_C \end{pmatrix}\right] \begin{pmatrix} \Psi_X \\ \Psi_C \end{pmatrix},
\end{equation*}
where the single particle polariton Hamiltonian $H_0$ is given by
\begin{equation*}
H_0=
\begin{pmatrix}
\omega_X-i\kappa_X/2 & \Omega_R/2 \\ \Omega_R/2 & \omega_C\left(-i\nabla\right)-i\kappa_C/2 \end{pmatrix},
\end{equation*}
and
\begin{equation*}
\omega_C\left(-i\nabla\right)=\omega_C(0)-\frac{\nabla^2}{2m_c}
\end{equation*}

\noindent
is the cavity dispersion, with the photon mass $m_C=
5\times10^{-5} m_0$ and $m_0$ the bare electron mass. For our
simulations we assumed a flat exciton dispersion relation
$\omega_X({\bf k})=\omega_X (0)$, set the exciton-photon detuning to
zero $\delta_{ex-ph}=\omega_X(0)-\omega_C(0)=0$ and set this energy
value as the reference of the zero energy. The parameters
$\Omega_R$, $\kappa_X$ and $\kappa_C$ are the Rabi frequency and the
excitonic and photonic decay rates respectively and have been
given values close to the usual
experimental ones: $\Omega_R=5.1$ meV, $\kappa_X= 0.05$ meV, and
$\kappa_C=0.08$ meV \cite{hivet_2014}. In this model polaritons are
injected into the cavity by four coherent and monochromatic laser
fields with pump intensity $f_p$ and Gaussian spatial
profiles with $\sigma_p$ of 20
$\mu$m: $F({\bf x})=\sum_{i=1}^4 f_pe^{i{\bf k_{p_i}}{\bf x}}
e^{-({\bf x}-{\bf x_i})^2/2\sigma^2_p}$. Where $\bf k_{p_i}$
are the four wave vectors of the four pumps that we fix to have
the same modulus $|{\bf k_{p_i}}|=|{\bf k}|$. In order 
to ensure that the phase of the polariton fluid is not
imposed by the laser pumps in the central region of the system we
set to zero the pump intensity outside of a $\sigma_{pin}=9\mu$m
radius circle. The exciton-exciton interaction
strength $g_X$ is set to one by rescaling both the cavity and
excitonic fields and the pump intensities. The numerical solution of
the GP equation is obtained over a two-dimensional grid (of
$512\times512$ points) in a box with sides of $150\times150\
\mu$m$^2$ using a fifth-order adaptive-step Runge-Kutta algorithm.
All the analysed quantities are taken when the system has reached a
steady state condition after a transient period of 200 ps.

To understand the role of the subsonic character of the fluid in the
annihilation of a vortex-antivortex pair we define the local fluid
velocity $v_f({\bf x})=\hbar |{\bf k}({\bf x})|/m_{LP}$, where $m_{LP}$ is
the lower polariton mass and ${\bf k}({\bf x})$ is the locally evaluated
derivative of the phase at the point ${\bf x}$. Moreover we define the quantity
$c_s({\bf x})=\sqrt{\hbar g_{LP} |\Psi_{LP}({\bf x})|^2/m_{LP}}$,
where $|\Psi_{LP}({\bf x})|^2$ and $g_{LP}$ are the local polariton
density and the coupling constant. Since in the local density
approximation $c_s({\bf x})$ corresponds to the speed of sound
defined in the case of high densities
\cite{wouters_spatial_2008,ciuti_quantum_2005}, we can take it as
definition of a {\it generalized local speed of sound} valid also
for low polariton densities and we define a generalized
Mach number:

\begin{equation}
\label{eqM}
M({\bf x})=\frac{v_f({\bf x})}{c_s({\bf x})}=\frac{\hbar |k({\bf x})|/m_{LP}}{\sqrt{\hbar g_{LP} |\Psi_{LP}({\bf x})|^2/m_{LP}}}.
\end{equation}

\noindent
This will allow to establish a direct correlation between the
subsonic or supersonic character of the fluid and the annihilation
of a V-AV pair. Since the subsonic character of the fluid is induced
by polariton-polariton interaction, this corresponds to studying the
role of polariton-polariton interactions in the annihilation of the pairs
and of the vortex-array. Although other techniques, like the study of
the vortex-antivortex correlation function \cite{liu_1992} or Reynolds-averaged
Navier-Stokes equations \cite{osborne_1895}, can be used to address
this problem, we chose to focus on the study of the Mach number
since it since it allows a simple and clear physical understanding
of the V-AV merging process.

\section{Results}\label{Results}
The mechanism lying beneath the annihilation of V-AV
pairs  and the role of polariton-polariton interactions in this
annihilation can be better highlighted by studying the
system behaviour as a function of the pump intensity and therefore
of the polariton density. We start by studying the system in the two
limiting cases of very low and very high pump intensities.

\begin{figure}
\includegraphics[width=8cm]{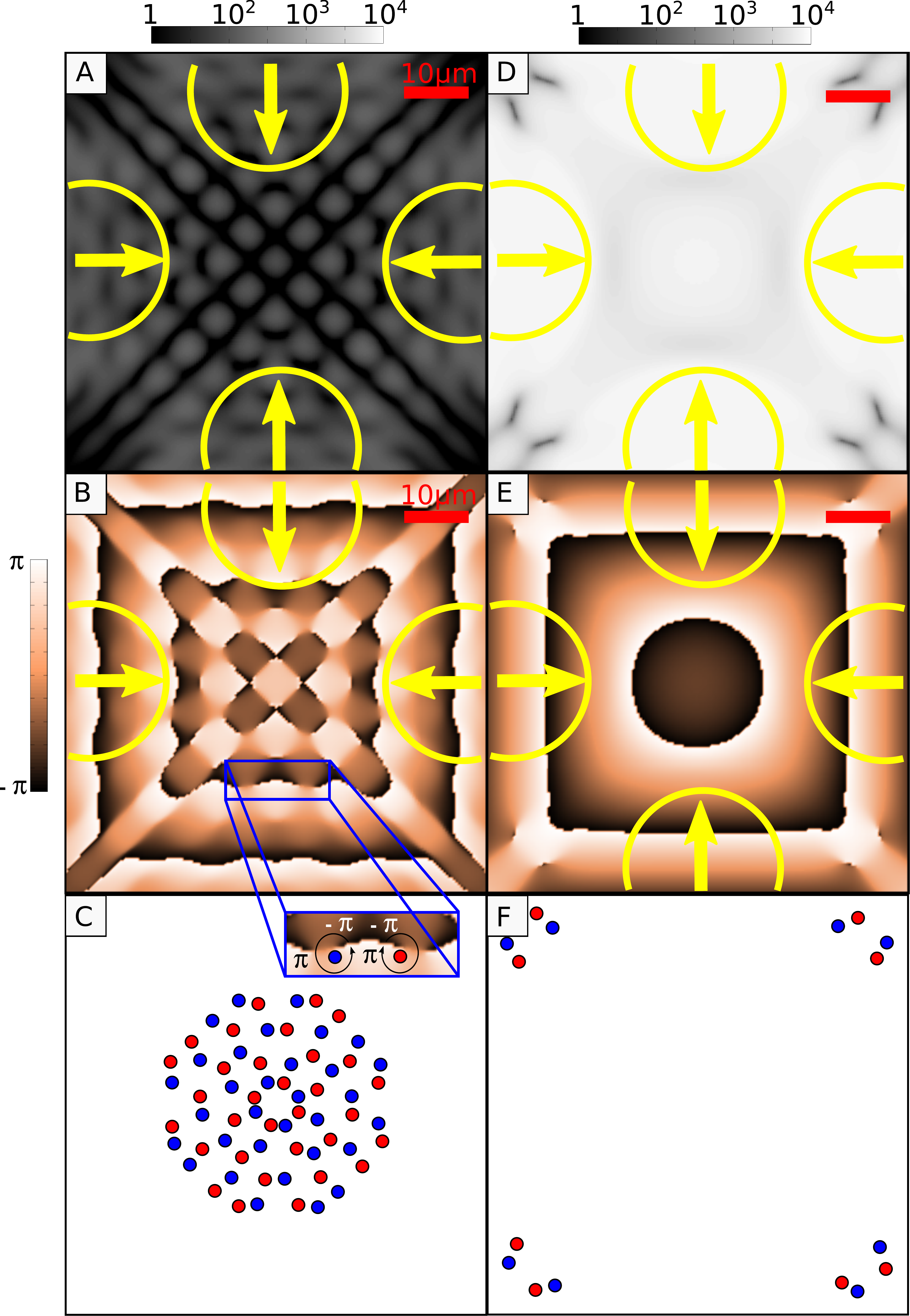}
\caption{\label{fig1} Numerical Real space emission intensity,
phase, and vortex distribution in the low and high density regimes.
The four polariton ensembles are
generated by pumps at $k_{pump}=0.7$ $\mu$m$^{-1}$ in different
directions (yellow arrows) with $\omega_{pump}=-2.25$ meV. A: Real
space image in the low density regime with pump intensity
$f_p=0.07f_{th}$, a square interference pattern with a unit cell of
about $9 \mu$m is formed. D: real space image in the high density
regime $f_p=3.33f_{th}$. B and E: Phases corresponding to A and D
showing an array of vortices and antivortices (B) and no phase
modulations (E). C and F: Vortices (red) and anti-vortices (blue)
distribution in real space corresponding to phase diagrams B to E.}
\end{figure}

In the low intensity case the polariton density lies on the lower
branch of the bistability curve everywhere in space and the system
behaviour is purely linear. In this regime (fig. \ref{fig1}A-C), the
formation of an array of vortices and antivortices is observed as in
\cite{hivet_2014,tosi_geometrically_2012,cristofolini_optical_2013}.
Here, like in \cite{hivet_2014}, the shape and size of the unit cell only
depend on the geometry of the pumping configuration and on the
angle of incidence of the laser beams. Since we use four pumps
with $|{\bf k}|=0.7 \mu$m$^{-1}$ the formed array has square unit cells
with unit cell size of approximately $9 \mu$m (fig. \ref{fig1}A-C).
This interference pattern generates an array of vortices (with a
clockwise phase winding from $-\pi$ to $\pi$) and antivortices (with
an anti-clockwise phase winding from $-\pi$ to $\pi$) that is
therefore due to purely linear mechanism (fig. \ref{fig1}C and inset
for the definition of vortices and antivortices).

In the opposite limit the pump  intensity is strong and
polariton-polariton interactions play a dominant role. This regime
is characterised by a polariton density
everywhere lying on the upper branch of the bistability curve. As it
can be seen in fig. \ref{fig1}D-F, in this regime the array of
vortices and antivortices has completely disappeared due to the
renormalisation of the lower-polariton branch.

From this, we can deduce that in the transition region between the
low and the high density regimes vortices and antivortices either
are expelled from the fluid or annihilate each-other. In order to
investigate this transition we vary the pump intensity around the
threshold value ($f_{th}$) at which V-AV pairs disappear. Note that
in this non-homogeneous system the density if different in the regions
within and outside the pumping spot, resulting in four threshold intensities:
two for increasing and two for decreasing pump intensities.
When the intensity of the four pumps increases the regions directly
pumped by the lasers jump from the lower to the upper branch of
the bistability curve (first threshold). At this point, since polaritons
have a finite lifetime, the central region between the four laser spots
is still in the lower part of the bistability curve. As the pumps intensity
is further increased also the central region eventually jumps from the
lower to the upper branch (second threshold). This threshold corresponds
to the intensity $f_{th}$ at which V-AV pairs disappear. Similarly one can
observe two threshold for decreasing pump intensity.

\begin{figure}
\includegraphics[width=8cm]{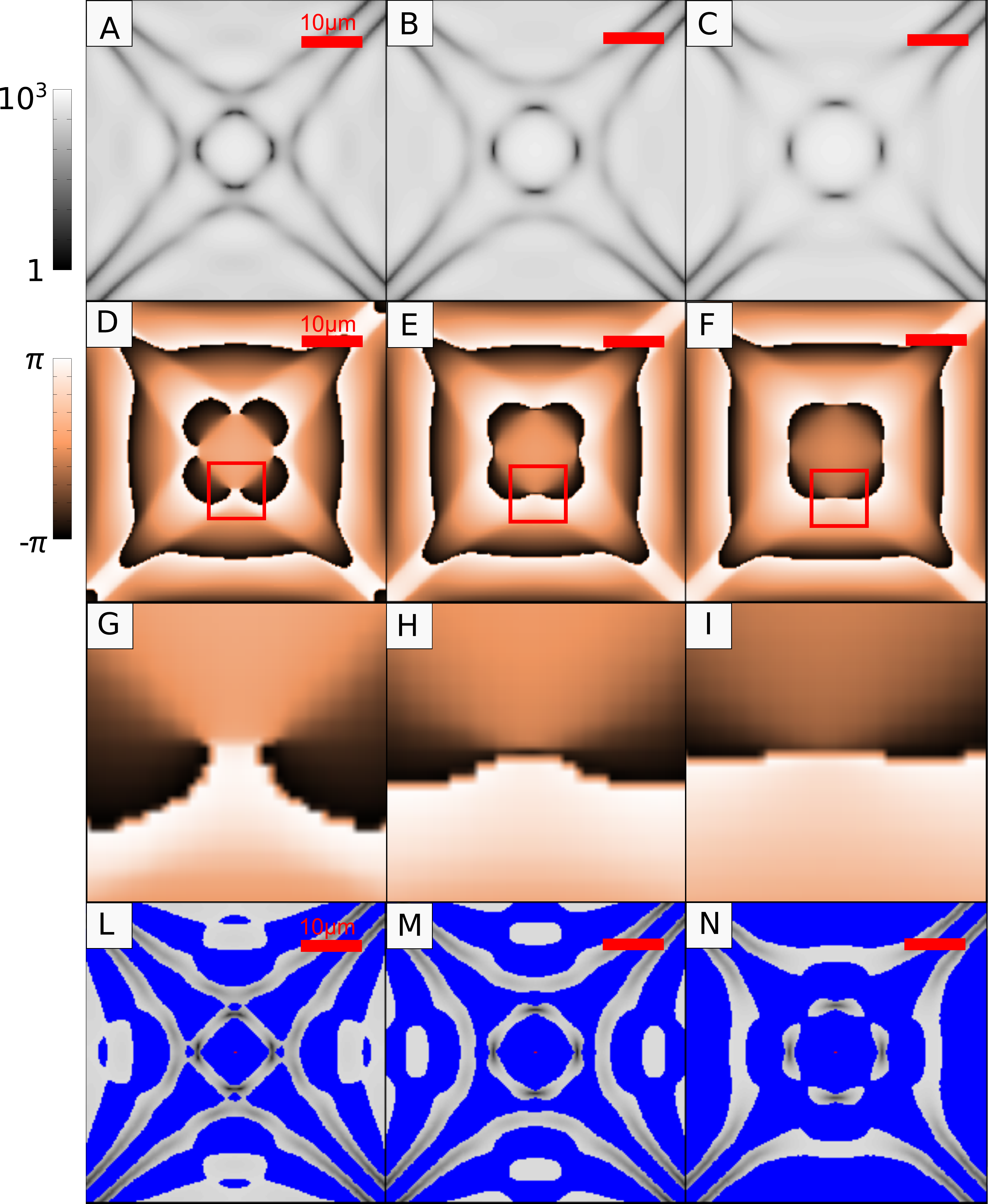}
\caption{\label{fig2} A-C: Numerical Real-space emission intensity
for three different pump intensities. D-F: real space phase-diagrams
corresponding to the pumping condition of A-C. G-I: Enlargement of
the red squared area in D-F. L-N: Mach charts corresponding to
panels A-C, where the fluid is supersonic the Mach chart show the
real-space emission intensity at that point, when the fluid is
subsonic the Mach chart is blue. A V-AV pair is clearly visible in
G, V and AV have almost merged in H, and no vortex pair is present
any more in I. The pump parameters are the same as in fig.
\ref{fig1} and the pump intensities (from left to right) are
$f_p=0.66f_{th}$, $0.8 f_{th}$, $1.0f_{th}$.}
\end{figure}

Figures \ref{fig2} A-C(D-F) represent the polariton distribution (phase) for increasing
pump intensity from just below to just above $f_{th}$. In fig. \ref{fig2}A
($f_p=0.66f_{th}$) some remaining of the interference pattern of fig. \ref{fig1}A is still
visible together with four dark segments surrounding the centre of
the image that correspond to four V-AV pairs. In this plot of the
intensity distribution each V-AV pair 
looks like a straight dark segment rather than
like two separated vortices because the core of the
vortex is extremely close to the core of the
antivortex. The fact that these four
dark segments correspond to V-AV pairs is confirmed by the phase
distribution of fig. \ref{fig2}D where the two phase rotations of
the vortex and of the antivortex are visible in correspondence of
each dark segment of fig. \ref{fig2}A. For the sake of clarity the
phase distribution of the V-AV pair delimited by the red square in
fig. \ref{fig2}D is also reported enlarged in fig. \ref{fig2}G. When
the pump intensity is increased (fig. \ref{fig2}B
\textbf{($f_p=0.8f_{th}$)} and fig. \ref{fig2}C
\textbf{($f_p=1.0f_{th}$)}) the vortex and the antivortex cores
get closer (i.e. the dark segments become
shorter) until the four V-AV pairs disappear. For
even higher pump intensities the density distribution becomes
homogeneous as in fig. \ref{fig1}D. Again this moving closer and
merging of the V-AV is confirmed by the corresponding phase
distributions (fig. \ref{fig2}D-F), and by the
corresponding zoom of the regions delimited by the red squares (fig.
\ref{fig2}G-I).

This detailed analysis of the density and phase distributions and of
the "on site" annihilation of V-AV pairs shows that the
disappearance of the vortex array is not due to the expulsion of the
vortices from the fluid. Therefore, the mechanism lying beneath the
V-AV annihilation cannot be ascribed to a simple renormalisation of
the lower polariton branch. First, the effect of the renormalisation is to
decrease the wave vector of the injected polaritons and correspondingly
to increase the size of the interference pattern of the array therefore
leading to the expulsion of vortices from the fluid rather than the observed
merging. Second, it can be seen in fig. \ref{fig2}C that when V-AV annihilation
takes place, part of the interference pattern is still visible. Third, vortices and antivortices
can exist in a fluid at rest, so the renormalization of the lower polariton
branch can not justify {\it per se} the disappearence of the array.

Moreover, since all the plots show the system steady state for a
given pump intensity, the fact that vortices coexist with antivortices
means that their position is the result of a tradeoff between vortex-vortex
and vortex-antivortex interactions, direction of the polariton flow,
and relative phase of the four pumps. In other words, this means that
vortices and antivortices are not completely free to move in the system
and therefore V-AV annihilation cannot be simply ascribed to V-AV
attraction. In fact, if vortices and antivortices are completely free to
move, due to their mutual attraction they must always annihilate, and
the only possible steady state must be completely free of vortex dislocations
independently from the intensity of the laser pumps. Finally, it is worth
noting that due to the choice of the continuous-wave resonant pumping
setup no exciton-reservoir is present in the system and all polaritons
have the same energy therefore no trapping mechanism can be advocated
to explain this annihilation as in \cite{cristofolini_optical_2013}.

\begin{figure}
\includegraphics[width=8cm]{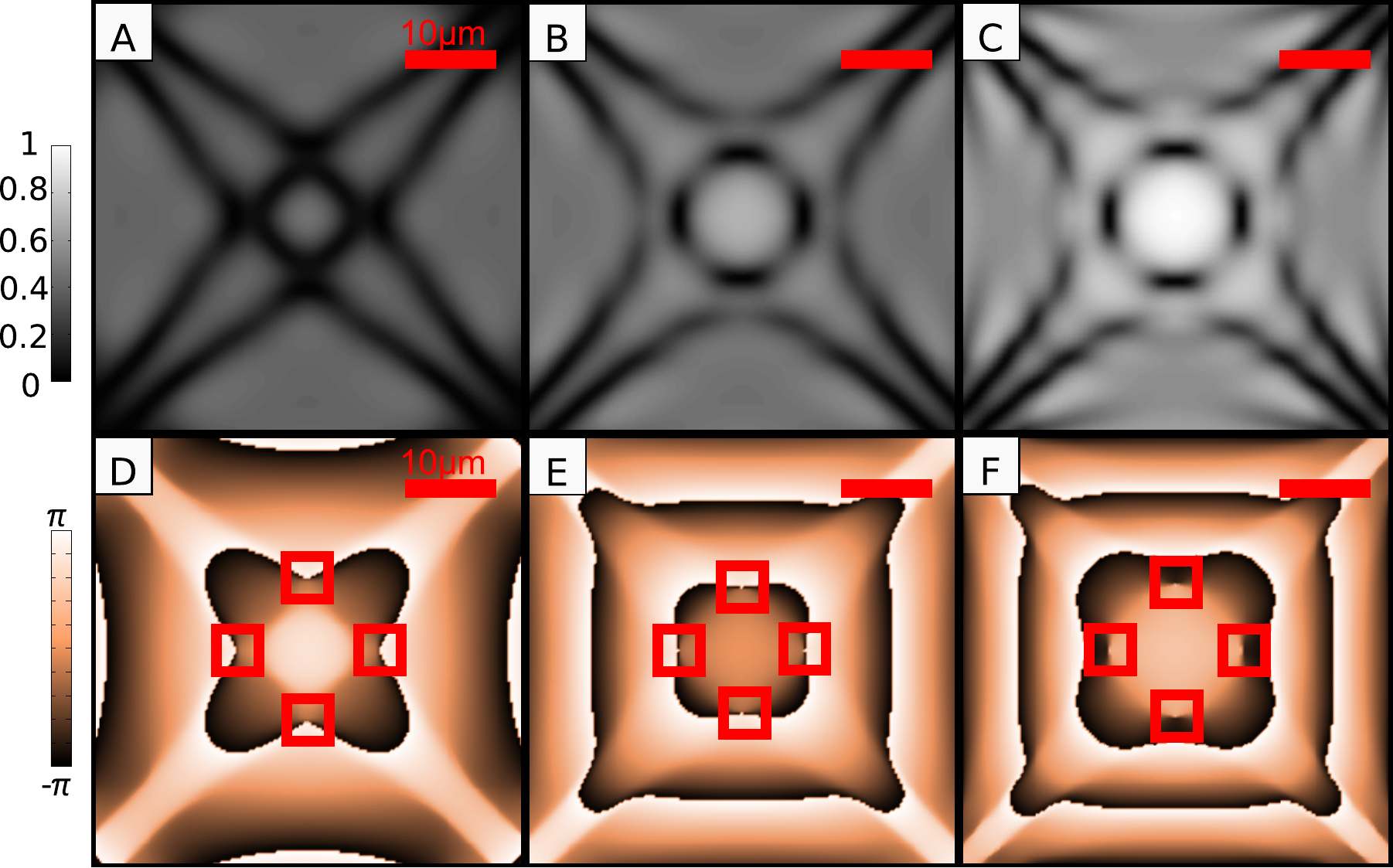}
\caption{\label{fig3} A-C: Numerical real-space emission intensity
(in linear greyscale) . D-F: Real space phase diagram (color scale)
corresponding to A-C. The red squares indicate the regions where
vortex and anti-vortex pairs have been annihilated. The plots
correspond to pumps with increasing in plane momenta from left to
right: $k_{pump}=0.5, 0.7$ and $0.9\ \mu$m$^{-1}$. In each case the
pump intensity is chosen in order to drive the system at the
threshold at which V-AV annihilation occurs ($f_p(B)=4.33f_p(A),
f_p(C)=8.66f_p(A)$).}
\end{figure}

To have a physical insight of the mechanism lying beneath V-AV annihilation it is
useful to study the Mach-number charts in fig. \ref{fig2}L-N where
blue regions correspond to a subsonic character of the fluid and
grey regions correspond to a supersonic character. Clearly as
the pump intensity is increased (from left to right) the polariton
density increases and the regions where the fluid is subsonic
becomes wider and wider. This spreading of the subsonic regions
rearranges the polariton flow therefore changing the steady
state position of vortices and antivortices until, when the pump
threshold $f_{th}$ is reached, vortices and antivortices annihilate. This
mechanism can also be understood observing that vortices cannot
 enter subsonic regions since they are intrinsically related to
regions where the fluid is supersonic, because at the centre of their
core the fluid density vanishes while the fluid velocity does not
and because a subsonic fluid tends to avoid strong phase modulations
\cite{landau_chapter_1986}. Since in our system the subsonic
regions surround the V-AV distribution, when the pump intensity
changes the polariton flows rearrange letting vortices overlap and
annihilate with antivortices. This is different from the case
of atomic Bose-Einstein condensates where a single vortex cannot be destroyed
by the superfluid character of a quantum fluid since the total angular
momentum of the system must be conserved. Here, polariton-polariton
interactions can affect the fluid distribution and make a vortex overlap
with an antivortex, so that V-AV pairs can be completely suppressed
in the fluid.

An additional proof of the correlation between the subsonic
character of the fluid and the disappearance of V-AV pairs is
given by the fact that higher sound velocities (i.e. higher
polariton densities) are needed to destroy the array when polaritons
have higher velocities. Figures \ref{fig3} A-C show the real space
distribution of the photonic field for three different increasing
velocities of the injected polaritons ($k_{pump}=0.5, 0.7$ and $0.9\
\mu$m$^{-1}$) and a pump intensity corresponding to the
threshold ($f_p=f_{th}$) where V-AV pairs disappear.
We find that the polariton density, and therefore the sound velocity,
at which V-AV pairs annihilate is higher when the velocity of the injected
polaritons is higher (see fig. \ref{fig3} D-F where no phase cut corresponding
to a V-AV pair can be observed in the four regions delimited by the
red squares). This confirms that polariton-polariton interactions, causing
the widening of the subsonic regions and the consequent rearrangement
of the polariton flows, induce the disappearance of V-AV pairs.

\section{Conclusions}\label{Conclusion}
We have investigated the washing out of vortex-antivortex lattices
in exciton polariton systems as a function of the polariton density.
Our detailed analysis shows that V-AV annihilation and the washing
out of the vortex lattice is due to polariton-polariton interactions that,
through the renormalization of the lower polariton branch, induce the
widening of the regions where the fluid is subsonic as the pump intensity
is increased. This widening induce a modification in the polariton flows
therefore changing the steady state position of vortices and antivortices
until a threshold pump intensity is reached at which V-AV merging takes place.
The complete washing out of the V-AV array can take place because
in our system has the number of vortices is equal to the number of
antivortices, i.e. the system has zero angular momentum. Our analysis
applied to a system with net angular momentum could open the way
to the study of vortex-vortex interactions.

\begin{acknowledgements}
We would like to thank C. Tejedor for the use of the computational
facilities of the Universidad Autonoma de Madrid and I. Carusotto and
F. M. Marchetti for useful discussions. This work has been partially
funded by the Quandyde project of the ANR France, by the POLATOM ESF
Research Network Program and by the CLERMONT4 Network Progam.
MHS from the EPSRC (grants EP/I028900/2 and EP/K003623/2).
A. B. is member of Institut Universitaire de France (IUF).
\end{acknowledgements}

\bibliography{bibli}{}

\begin{thebibliography}{10}

\bibitem{desyatnikov_optical_2005}
A.~S. Desyatnikov, Y.~S. Kivshar, and L.~Torner, ``Optical vortices and vortex
  solitons,'' in {\em Progress in Optics} (E.~Wolf, ed.), vol.~47,
  pp.~291--391, Elsevier, Amsterdam, 2005.

\bibitem{essmann_direct_1967}
U.~Essmann and H.~Träuble, ``The direct observation of individual flux lines in
  type {II} superconductors,'' {\em Physics Letters A}, vol.~24, pp.~526--527,
  May 1967.

\bibitem{yarmchuk_observation_1979}
E.~J. Yarmchuk, M.~J.~V. Gordon, and R.~E. Packard, ``Observation of stationary
  vortex arrays in rotating superfluid helium,'' {\em Physical Review Letters},
  vol.~43, pp.~214--217, July 1979.

\bibitem{scheuer_optical_1999}
J.~Scheuer and M.~Orenstein, ``Optical vortices crystals: Spontaneous
  generation in nonlinear semiconductor microcavities,'' {\em Science},
  vol.~285, pp.~230--233, Sept. 1999.

\bibitem{madison_vortex_2000}
K.~W. Madison, F.~Chevy, W.~Wohlleben, and J.~Dalibard, ``Vortex formation in a
  stirred bose-einstein condensate,'' {\em Physical Review Letters}, vol.~84,
  pp.~806--809, Jan. 2000.

\bibitem{denschlag_generating_2000}
J.~Denschlag, J.~E. Simsarian, D.~L. Feder, C.~W. Clark, L.~A. Collins,
  J.~Cubizolles, L.~Deng, E.~W. Hagley, K.~Helmerson, W.~P. Reinhardt, S.~L.
  Rolston, B.~I. Schneider, and W.~D. Phillips, ``Generating solitons by phase
  engineering of a bose-einstein condensate,'' {\em Science}, vol.~287,
  pp.~97--101, July 2000.

\bibitem{khaykovich_formation_2002}
L.~Khaykovich, F.~Schreck, G.~Ferrari, T.~Bourdel, J.~Cubizolles, L.~D. Carr,
  Y.~Castin, and C.~Salomon, ``Formation of a matter-wave bright soliton,''
  {\em Science}, vol.~296, pp.~1290--1293, May 2002.

\bibitem{lagoudakis_quantized_2008}
K.~G. Lagoudakis, M.~Wouters, M.~Richard, A.~Baas, I.~Carusotto, R.~André,
  L.~S. Dang, and B.~Deveaud-Plédran, ``Quantized vortices in an
  exciton-polariton condensate,'' {\em Nature Physics}, vol.~4, pp.~706--710,
  Sept. 2008.

\bibitem{lagoudakis_observation_2009}
K.~G. Lagoudakis, T.~Ostatnický, A.~V. Kavokin, Y.~G. Rubo, R.~André, and
  B.~Deveaud-Plédran, ``Observation of half-quantum vortices in an
  exciton-polariton condensate,'' {\em Science}, vol.~326, pp.~974--976, Nov.
  2009.
\newblock {PMID:} 19965506.

\bibitem{flayac_topological_2010}
H.~Flayac, I.~A. Shelykh, D.~D. Solnyshkov, and G.~Malpuech, ``Topological
  stability of the half-vortices in spinor exciton-polariton condensates,''
  {\em Physical Review B}, vol.~81, p.~045318, Jan. 2010.

\bibitem{roumpos_single_2011}
G.~Roumpos, M.~D. Fraser, A.~Löffler, S.~Höfling, A.~Forchel, and Y.~Yamamoto,
  ``Single vortex-antivortex pair in an exciton-polariton condensate,'' {\em
  Nature Physics}, vol.~7, pp.~129--133, Feb. 2011.

\bibitem{nardin_hydrodynamic_2011}
G.~Nardin, G.~Grosso, Y.~Léger, B.~Pi\c{e}tka, F.~Morier-Genoud, and
  B.~Deveaud-Plédran, ``Hydrodynamic nucleation of quantized vortex pairs in a
  polariton quantum fluid,'' {\em Nature Physics}, vol.~7, pp.~635--641, Aug.
  2011.

\bibitem{tosi_onset_2011}
G.~Tosi, F.~M. Marchetti, D.~Sanvitto, C.~Antón, M.~H. Szymanska, A.~Berceanu,
  C.~Tejedor, L.~Marrucci, A.~Lemaître, J.~Bloch, and L.~Viña, ``Onset and
  dynamics of vortex-antivortex pairs in polariton optical parametric
  oscillator superfluids,'' {\em Physical Review Letters}, vol.~107, p.~036401,
  July 2011.

\bibitem{sanvitto_all-optical_2011}
D.~Sanvitto, S.~Pigeon, A.~Amo, D.~Ballarini, M.~D. Giorgi, I.~Carusotto,
  R.~Hivet, F.~Pisanello, V.~G. Sala, P.~S.~S. Guimaraes, R.~Houdré,
  E.~Giacobino, C.~Ciuti, A.~Bramati, and G.~Gigli, ``All-optical control of
  the quantum flow of a polariton condensate,'' {\em Nature Photonics}, vol.~5,
  no.~10, pp.~610--614, 2011.

\bibitem{gorbach_vortex_2010}
A.~V. Gorbach, R.~Hartley, and D.~V. Skryabin, ``Vortex lattices in coherently
  pumped polariton microcavities,'' {\em Physical Review Letters}, vol.~104,
  p.~213903, May 2010.

\bibitem{keeling_spontaneous_2008}
J.~Keeling and N.~G. Berloff, ``Spontaneous rotating vortex lattices in a
  pumped decaying condensate,'' {\em Physical Review Letters}, vol.~100,
  p.~250401, June 2008.

\bibitem{tosi_geometrically_2012}
G.~Tosi, G.~Christmann, N.~G. Berloff, P.~Tsotsis, T.~Gao, Z.~Hatzopoulos,
  P.~G. Savvidis, and J.~J. Baumberg, ``Geometrically locked vortex lattices in
  semiconductor quantum fluids,'' {\em Nature Communications}, vol.~3, p.~1243,
  Dec. 2012.

\bibitem{cristofolini_optical_2013}
P.~Cristofolini, A.~Dreismann, G.~Christmann, G.~Franchetti, N.~G. Berloff,
  P.~Tsotsis, Z.~Hatzopoulos, P.~G. Savvidis, and J.~J. Baumberg, ``Optical
  superfluid phase transitions and trapping of polariton condensates,'' {\em
  Physical Review Letters}, vol.~110, p.~186403, May 2013.

\bibitem{kusudo_stochastic_2012}
K.~Kusudo, N.~Y. Kim, A.~Loeffler, S.~Hoefling, A.~Forchel, and Y.~Yamamoto,
  ``Stochastic formation of polariton condensates in two degenerate orbital
  states,'' {\em {arXiv:1211.3833}}, Nov. 2012.

\bibitem{liew_generation_2008}
T.~C.~H. Liew, Y.~G. Rubo, and A.~V. Kavokin, ``Generation and dynamics of
  vortex lattices in coherent exciton-polariton fields,'' {\em Physical Review
  Letters}, vol.~101, p.~187401, Oct. 2008.

\bibitem{hivet_2014}
R.~Hivet, E.~Cancellieri, T.~Boulier, D.~Ballarini, D.~Sanvitto, F.~M.
  Marchetti, M.~H. Szymanska, C.~Ciuti, E.~Giacobino, and A.~Bramati,
  ``Interaction-shaped vortex-antivortex lattices in polariton fluids,'' {\em
  Phys. Rev. B}, vol.~89, p.~134501, Apr 2014.

\bibitem{carusotto_quantum_2013}
I.~Carusotto and C.~Ciuti, ``Quantum fluids of light,'' {\em Reviews of Modern
  Physics}, vol.~85, pp.~299--366, Feb. 2013.

\bibitem{wouters_spatial_2008}
M.~Wouters, I.~Carusotto, and C.~Ciuti, ``Spatial and spectral shape of
  inhomogeneous nonequilibrium exciton-polariton condensates,'' {\em Physical
  Review B}, vol.~77, p.~115340, Mar. 2008.

\bibitem{ciuti_quantum_2005}
C.~Ciuti and I.~Carusotto, ``Quantum fluid effects and parametric instabilities
  in microcavities,'' {\em physica status solidi (b)}, vol.~242, no.~11,
  pp.~2224--2245, 2005.

\bibitem{liu_1992}
F.~Liu and G.~F. Mazenko, ``Defect-defect correlation in the dynamics of
  first-order phase transitions,'' {\em Phys. Rev. B}, vol.~46, pp.~5963--5971,
  Sep 1992.

\bibitem{osborne_1895}
R.~Osborne, ``On the dynamical theory of incompressible viscous fluids and the
  determination of the criterion,'' {\em Philosophical Transactions of the
  Royal Society of London. A}, vol.~186, pp.~123--164, 1895.

\bibitem{landau_chapter_1986}
L.~D. Landau and E.~M. Lifshitz, ``chapter 6 : Superfluidity,'' in {\em
  Statistical Physics part 1 (2nd edition)}, Course of Theoretical Physics,
  Pergamon Press, Oxford, 1986.

\end{thebibliography}
\bibliographystyle{ieeetr}

\end{document}